\documentclass[11pt, a4paper]{article}

\usepackage{amsmath}
\usepackage{yhmath}
\usepackage{amsfonts}
\usepackage{amssymb}
\usepackage{graphicx, rotating}
\usepackage{epstopdf}
\usepackage{epsfig}
\usepackage{latexsym}
\usepackage{graphicx}
\usepackage{color}
\usepackage{amsmath,bm,amssymb}
\usepackage{cite}
\usepackage{slashed}
\usepackage{hyperref}
\usepackage{dsfont} 
\hypersetup{colorlinks, citecolor=bluscuro, linkcolor=black, urlcolor=bluscuro}
\definecolor{rossos}{cmyk}{0,1,1,0.55}
\definecolor{bluscuro}{rgb}{0.15, 0.2, .85}
\definecolor{bluchiaro}{cmyk}{1,.3,0.,0.1}

\usepackage{mathtools}

\usepackage{amsmath}
\usepackage{amsfonts}
\usepackage{amssymb}
\usepackage{bbold}
\usepackage{mathrsfs}
\usepackage{latexsym} 
\usepackage{cancel}
\usepackage{bm}
\usepackage{graphicx, rotating}
\usepackage{epstopdf}
\usepackage{epsfig}
\usepackage{latexsym}
\usepackage{color}
\usepackage[dvipsnames]{xcolor}
\usepackage{cite}
\usepackage{slashed}
\usepackage{comment}
\usepackage[utf8]{inputenc}
\usepackage{soul}
\usepackage{multirow}
\usepackage{graphics,subfigure}
\usepackage[dvipsnames]{xcolor}
\usepackage{physics}
\usepackage[]{breqn} 

\definecolor{bluscuro}{rgb}{0.15, 0.2, .85}

\newcommand{\eq}[1]{Eq.~(\ref{#1})}

\newcommand{\nn}{\nonumber}

\newcommand{\be}{\begin{equation}}
\newcommand{\ee}{\end{equation}}
\newcommand{\bea}{\begin{eqnarray}}
\newcommand{\eea}{\end{eqnarray}}

\newcommand\blfootnote[1]{%
  \begingroup
  \renewcommand\thefootnote{}\footnote{#1}%
  \addtocounter{footnote}{-1}%
  \endgroup
}

 \def\bea{\begin{eqnarray}}
  \def\eea{\end{eqnarray}}

	\def \beq {\begin{equation}}
	\def \eeq {\end{equation}}
	\def \ba {\begin{array}}
	\def \ea {\end{array}}

	\def \ecart {\noalign{\medskip}}
	\def \dis {\displaystyle}

\newcommand{\orcid}{\includegraphics{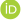}}
\newcommand{\orcidlink}[1]{\href{https://orcid.org/#1}{{\orcid}}}

\def \Booo{a_{000}}
\def \Bopo{a_{010}}
\def \Boop{a_{001}}
\def \Bopp{a_{011}}
\def \Bpoo{a_{100}}
\def \Bppo{a_{110}}
\def \Bpop{a_{101}}
\def \Bppp{a_{111}}
\def \Bijo{a_{ ij0}}
\def \Bijp{a_{ ij1}}

\begin{document}

\begin{titlepage}
\begin{flushright}
IFT-UAM/CSIC-23-90
\end{flushright}
\begin{center} ~~\\
\vspace{0.5cm} 
\Large {\bf\Large 
Entanglement and entropy in multipartite systems: 
\\ a useful approach
} 
\vspace*{1.5cm}

\normalsize{
{\bf 
A. Bernal
\orcidlink{0000-0003-3371-5320}
\blfootnote{
alexander.bernal@csic.es},
J.~A. Casas
\orcidlink{0000-0001-5538-1398}
\blfootnote{j.alberto.casas@gmail.com
} and
J.~M. Moreno
\orcidlink{0000-0002-2941-0690}
 } \\
 
\smallskip  \medskip
{\it Instituto de F\'\i sica Te\'orica, IFT-UAM/CSIC,}\\
\it{Universidad Aut\'onoma de Madrid, Cantoblanco, 28049 Madrid, Spain}}

\medskip

\vskip0.6in 

\end{center}

\centerline{ \large\bf Abstract }
\vspace{.5cm}

Quantum entanglement and quantum entropy are crucial concepts in the study of multipartite quantum systems. In this work we show how the notion of concurrence vector, re-expressed in a particularly useful form, provides new insights and computational tools for the analysis of both. 
In particular, using this approach for a general multipartite pure state, one can easily prove known relations in an easy way and to build up new relations between the concurrences associated with the different bipartitions. The approach is also useful to derive sufficient conditions for genuine entanglement in generic multipartite systems that are computable in polynomial time. From 
an entropy-of-entanglement perspective, the approach is powerful to prove properties
of the Tsallis-$2$ entropy, such as the subadditivity, and to derive new ones, e.g. a modified version of the strong subadditivity which is always fulfilled; thanks to the purification theorem these results hold for any multipartite state, whether pure or mixed.

\vspace*{2mm}
\end{titlepage}


\tableofcontents

\section{Introduction}
\label{sec:Intro}

Entanglement is a key notion in quantum mechanics and in the highly developing field of quantum information. For bipartite systems, ${\cal H} = {\cal H}_{A}\otimes {\cal H}_{B}$, there are rather simply criteria to detect and quantify the entanglement of pure states. In particular, if the state is entangled the reduced density matrix,  $\rho_A=\tr_B \rho$, corresponds to a mixed state, with $\tr \rho_A^2<1$. Accordingly, the value of the concurrence, 
$C^2\propto \left(1-\tr\rho_{A}^2\right)$ or the von Neumann entropy, $S=-\tr \rho_A\ln \rho_A$ are standard measures of  entanglement.

Things get more involved for multipartite Hilbert spaces, ${\cal H} = {\cal H}_{1}\otimes {\cal H}_{2}\otimes
\cdots
\otimes {\cal H}_{N}$. In that case there are $2^{N-1}-1$ inequivalent bipartitions. If the system is entangled with respect to all of them, it is called a genuine multipartite entangled state. Apart from its theoretical importance, genuine entanglement is relevant for applications such as quantum cryptography \cite{Ekert:1991zz}, quantum computation \cite{PhysRevLett.86.5188}, quantum teleportation \cite{PhysRevLett.70.1895}, high-precision metrology \cite{Giovannetti_2004}, spin chains\cite{PhysRevA.72.014301} and even biological systems \cite{Sarovar_2010}. The structure of genuine entanglement is quite intricate and its complexity grows exponentially with the number of subsystems. Actually, the problem of classifying and quantifying the genuine entanglement of general multipartite systems is still open (see  \cite{Guhne:20091,eltschka2014, PhysRevA.67.012108, Guhne_2010, PhysRevLett.111.030501, Ma_2011, PhysRevA.102.062424, Jin_2023} for relevant work in this subject).

In this work we explore the power of the ``concurrence vector" notion \cite{PhysRevA.64.052304,PhysRevA.62.012311,Akhtarshenas_2005,Li_2008} 
to extract powerful information about the entanglement structure of a general multipartite pure state. In section \ref{Preliminaries} we review the concept of concurrence vector and introduce a simple expression which turns out to be quite useful to establish connections between the entanglement of different bipartitions (section \ref{connections}). In section \ref{sec: triangular} we show that this formalism allows a simple derivation and an intuitive, geometrical, understanding of several triangular inequalities presented in the literature. Also, it allows to obtain new inequalities and generalize results that were only proven for a system of three qubits (the complete proof of the latter is given in the Appendix). In section \ref{sec:entropy} we explore the concurrence from an entropy-of-entanglement perspective, showing that the concurrence vector framework allows to easily prove the subadditivity condition for the Tsallis-2 entropy (directly related to the concurrence), as well as the violation of the strong subadditivity condition. In this sense we propose a modified version of the strong subadditivity condition which is always fulfilled, and derive other new relations. Thanks to the purification theorem these results are completely general for multipartite states, whether pure or mixed.
In section \ref{sec: suf_conds} we exploit the concurrence vector approach to formulate very simple sufficient conditions for genuine entanglement, which are computable in polynomial time. Finally, the conclusions are presented in section \ref{conclusions}.

\section{Preliminaries}\label{Preliminaries}
\subsection{Notation}

We start with a multipartite Hilbert space
\be
{\cal H} = {\cal H}_{1}\otimes {\cal H}_{2}\otimes
\cdots
\otimes {\cal H}_{N}\,,
\label{Hilbert}
\ee
using indices $i,j,k,\cdots$ to design the basis vectors of ${\cal H_{\rm 1}, H_{\rm 2} , H_{\rm 3}  ...\ }$. Each index takes as many values as the dimension of the corresponding Hilbert space.
We are not imposing any restriction here, i.e. each Hilbert space may have a different (finite) dimension.

Consider a bipartition ${\cal H} = {\cal H}_{ A}\otimes {\cal H}_{\wideparen{A}}$, and denote $I$ the subset of indices 
associated with ${\cal H}_{A}$ and $\wideparen{I}$ 
the complementary set of indices, associated with ${\cal H}_{\wideparen{A}}$  (in general we will denote collective indices with capital letters). Suppose now that the system is in a normalized pure state 
\be
|\psi\rangle = \sum_{i,j,k...}a_{ijk...} |i\rangle|j\rangle|k\rangle\cdots\,,
\label{psi}
\ee
with density matrix $\rho=|\psi\rangle\langle\psi|$.
Then, the entanglement of this state with respect to the previous bipartition is often
quantified by the concurrence \cite{PhysRevLett.78.5022,PhysRevLett.80.2245,PhysRevA.62.032307,PhysRevA.64.042315, PhysRevA.67.012307,albeverio2001,PhysRevA.59.141, PhysRevA.67.054305,Krynytskyi:2018rlc}, 
\be
C^2_{I|\wideparen{I}}=
2
\left(1-\tr\rho_{A}^2\right)\,,
\label{concurrence}
\ee
where $\rho_{A}$ is the reduced density matrix, i.e. after tracing in $\wideparen{I}$. 
(Other normalizations of $C^2_{I|\wideparen{I}}$ , as well as alternative equivalent definitions, can be found in the literature.)
The important point is that the state is biseparable if and only if $C^2_{I|\wideparen{I}}=0$.

\subsection{The concurrence vector}

Consider again the previous bipartition
${\cal H} = {\cal H}_{ A}\otimes {\cal H}_{\wideparen{A}}$. Clearly, the state (\ref{psi}) is separable as long as it can be written as 
\be
|\psi\rangle = \left(\sum_I 
\alpha_{I} |I\rangle\right)\otimes \left(\sum_{\wideparen{I}}\beta_{\wideparen{I}}\ |\wideparen{I}\rangle\right)\,,
\ee
so that
\be
a_{ijk...}=a_{I\wideparen{I}}=\alpha_I \beta_{\wideparen{I}}\,.
\ee
Then, if we view $a_{I\wideparen{I}}$ as a matrix of coefficients, for a separable state it has rank=1, which means that all its $2\times2$ minors are vanishing. Denoting by
\be
\left[a\right]_{\{I_1I_2\}\{\wideparen{I}_1\wideparen{I}_2\}}=a_{I_1\wideparen{I}_1}a_{I_2\wideparen{I}_2}-a_{I_1\wideparen{I}_2}a_{I_2\wideparen{I}_1}
\ee
the minor corresponding to the rows $I_1, I_2$ and columns $\wideparen{I}_1, \wideparen{I}_2$, the state is biseparable {\em iff}
\be
\sum_{I_1<I_2, \wideparen{I}_1<\wideparen{I}_2}\left|\left[a\right]_{\{I_1I_2\}\{\wideparen{I}_1\wideparen{I}_2\}}\right|^2
=\frac{1}{4}
\sum_{I_1,I_2, \wideparen{I}_1,\wideparen{I}_2}\left|\left[a\right]_{\{I_1I_2\}\{\wideparen{I}_1\wideparen{I}_2\}}\right|^2=0\,.
\label{sumofminors}
\ee
Here $I_1, I_2$ ($\wideparen{I}_1, \wideparen{I}_2$) 
run over all possible values of $I$ ($\wideparen{I}$).
The latter expression is actually equivalent to the concurrence definition (\ref{concurrence}). To check this, notice that
\bea
\frac{1}{4}
\sum_{I_1,I_2, \wideparen{I}_1,\wideparen{I}_2}\left|\left[a\right]_{\{I_1I_2\}\{\wideparen{I}_1\wideparen{I}_2\}}\right|^2=
\frac{1}{4}
\sum_{I_1,I_2, \wideparen{I}_1,\wideparen{I}_2}\left[a\right]_{\{I_1I_2\}\{\wideparen{I}_1\wideparen{I}_2\}}\left[a^\dagger\right]_{\{\wideparen{I}_1\wideparen{I}_2\}\{I_1I_2\}}
={\frac{1}{2}}\sum_{I_1,I_2}\left[a a^\dagger\right]_{\{I_1I_2\}\{I_1I_2\}}\,,
\eea
where the last expression is just the sum of the principal minors of the reduced density matrix $aa^\dagger=\rho_{A}$. On the other hand, by Cauchy-Binet theorem,
\be
\frac{1}{2}
\sum_{I_1,I_2}\left[\rho_A\right]_{\{I_1I_2\}\{I_1I_2\}}=\frac{1}{2}\sum_{I_1\neq I_2}\lambda_{I_1}\lambda_{I_2}=\frac{1}{2}(1-\sum\lambda_{I}^2)=
\frac{1}{2}(1-\tr\rho_{A}^2)\,,
\ee
where $\lambda_I$ are the eigenvalues of $\rho_A$ and we have used $\sum\lambda_I=1$. In summary, the concurrence definition (\ref{concurrence}) is equivalent to 
\be
C^2_{I|\wideparen{I}}=
\sum_{I_1,I_2, \wideparen{I}_1,\wideparen{I}_2}\left|\left[a\right]_{\{I_1I_2\}\{\wideparen{I}_1\wideparen{I}_2\}}\right|^2=\|\vec{C}_{I|\wideparen{I}}\|^2\,,
\ee
where the ``concurrence vector" is the ordered list of all the minors of  
the matrix of coefficients $a_{I\wideparen{I}}$\ :
\be
\vec{C}_{I|\wideparen{I}}=
\{\left[a\right]_{\{I_1I_2\}\{\wideparen{I}_1\wideparen{I}_2\}} \}
=\{ a_{I_1\wideparen{I}_1}a_{I_2\wideparen{I}_2}-a_{I_1\wideparen{I}_2}a_{I_2\wideparen{I}_1} \}\,.
\label{convec}
\ee
Note that, denoting $D={\rm dim}\  {\cal H}$,
the concurrence vector has length $D^2$, since $a_{I\wideparen{I}}$
has $D$ entries. The previous notion of concurrence vector has been already considered in the literature 
\cite{PhysRevA.64.052304,PhysRevA.62.012311,Akhtarshenas_2005,Li_2008}.

\subsection{A useful expression}

Let us express the concurrence vector (\ref{convec}) in a more useful way. First define the (dimension $D^2$) vector $\vec{A}$, with components
\be
A_{i_1 j_1 k_1...;\ i_2 j_2 k_2...}=(a_{i_1 j_1 k_1...})\,  (a_{i_2j_2 k_2...})
\label{def_A}
\ee
for all possible values of the indices. Notice that $A_{i_1 j_1 k_1...;\, i_2 j_2 k_2...}$ are simply the coefficients of the state $|\psi\rangle \otimes |\psi\rangle \in {\cal H}\otimes{\cal H}$. If 
all the coefficients are real the entries of $\vec{A}$ coincide with those of the density matrix, but this is not the general case.
Under a permutation of the $i-$index, say $P_i$, associated with the Hilbert space ${\cal H}_1$, $\vec{A}$ changes as\footnote{If we view $A_{i_1 j_1 k_1,...;\ i_2,j_2 k_2,...}$  as an $D\times D$ matrix (analogous to the density matrix associated with $|\psi\rangle \langle\psi|$), then this permutation is equivalent to perform a partial transpose in the $i-$index.} 
\be
P_i \vec{A}= \{(a_{i_2 j_1 k_1...})\ (a_{i_1j_2 k_2...})\}\,.
\ee
This is simply a certain re-ordering of the components of $\vec{A}$.
Note that $P_i$ are Hermitian linear operators satisfying $P_i^2=1$ and $[P_i, P_j]=0$. Let us now consider an ``elementary bipartition", i.e. one between one of the initial parties and the rest, say ${\cal H}_1\otimes ({\cal H}_2\otimes {\cal H}_3\otimes\cdots)$. Then the concurrence vector (\ref{convec}) (with $I=i$) reads
\be
\vec{C}_{i|\wideparen{i}}=
\{ a_{i_1\wideparen{i}_1}a_{i_2\wideparen{i}_2}-a_{i_1\wideparen{i}_2}a_{i_2\wideparen{i}_1} \} = 
(\mathds{1}-P_i)\vec{A}\,.
\label{convec2}
\ee
Similarly, the concurrence vector (\ref{convec}) associated to the bipartition 
${\cal H}= ({\cal H}_1\otimes {\cal H}_2)\otimes ({\cal H}_3\otimes\cdots)$ 
reads
\be
\vec{C}_{ij|\wideparen{ij}}=
\{ a_{i_1j_1\wideparen{i_1j_1}}\ a_{i_2j_2\wideparen{i_2j_2}}\ -\  a_{i_1j_1\wideparen{i_2 j_2}}\ a_{i_2j_2\wideparen{i_1j_1}} \} = 
(\mathds{1}-P_iP_j)\vec{A}\,.
\label{convec3}
\ee
These expressions are straightforwardly extended for other bipartitions. In general, for a bipartition $I|\wideparen{I}$, with $I={i,j,...,m}$ (or any other subset of indices) the concurrence vector reads
\be
\vec{C}_{I|\wideparen{I}}=\vec{C}_{\wideparen{I}|I}=(\mathds{1}-P_I)\vec{A}\ \equiv
(\mathds{1}-P_iP_j\cdots P_m)\vec{A}\,.
\label{convec4}
\ee
Mathematically, the elementary permutations, $P_i, P_j, \dots$ are the generators of the group of all permutations of 
indices, $\{P_I\}$, which is a commutative $Z_2^N$ group. 
Note that $P_I\vec{A}=P_{\wideparen{I}}\vec{A}$
since $\vec{A}$
is obviously symmetric under the interchange of all indices.
So, the group contains $2^{N-1}$ inequivalent permutations, 
corresponding to the possible bipartitions of ${\cal H}$. This includes the trivial bipartition ${\cal H} = {\cal H}_{ A}\otimes {\emptyset}$, corresponding to the identity of the group. So, there are in fact $2^{N-1}-1$ non-trivial permutations and bipartitions. For each one, the associated concurrence vector is given by (\ref{convec4}). Note also that $(\mathds{1}-P_I)$ are ``projectors", satisfying $(\mathds{1}-P_I)^2=2(\mathds{1}-P_I)$.

Next we will exploit the notion of concurrence vector and its latter expression (\ref{convec4}) to derive some direct consequences.

\section{Connection between the entanglements of different bipartitions}\label{connections}

The idea of concurrence vector allows to derive the entanglement of any of the $2^{N-1}-1$ different bipartitions from the entanglement of the $N$ elementary bipartitions. In other words, we can use the concurrence vectors of the latter as building blocks to construct any other concurrence.

For example, the concurrence of two parts with respect to the rest, say $\vec{C}_{ij|\wideparen{ij}}$, is related to the elementary ones, $\vec{C}_{i|\wideparen{i}}\ $, $\vec{C}_{j|\wideparen{j}}\,$, by
\bea
\vec{C}_{ij|\wideparen{ij}}=(\mathds{1}-P_iP_j)\vec{A}= P_j\Big((\mathds{1}-P_i)-(\mathds{1}-P_j)\Big)\vec{A} = 
P_j\left(\vec{C}_{i|\wideparen{i}}-\vec{C}_{j|\wideparen{j}}\right)\,,
\label{2to1}
\eea
or equivalently 
\bea
\vec{C}_{ij|\wideparen{ij}}=(\mathds{1}-P_iP_j)\vec{A}= \Big((\mathds{1}-P_i)+P_i(\mathds{1}-P_j)\Big)\vec{A} = 
\vec{C}_{i|\wideparen{i}}+P_i\vec{C}_{j|\wideparen{j}}\,.
\label{2to1Alt}
\eea
More generically, the concurrence vector, $\vec{C}_{I|\wideparen{I}}$, of an arbitrary bipartition, say $I=\{i,j,...., m,n\}$, can be expressed as
\bea
\vec{C}_{I|\wideparen{I}} &=& (\mathds{1}-P_iP_j\cdots P_n) \vec{A}
\nonumber\\
&=& 
\big[ \left(\mathds{1}-P_i\right) + \left(P_i-P_iP_j\right)+\cdots
+ \left(P_iP_j...P_m- P_iP_j...P_mP_n\right)
\big]\vec{A}
\nonumber\\
&=& \vec{C}_{i|\wideparen{i}}+P_i \vec{C}_{j|\wideparen{j}}+P_i P_j\vec{C}_{k|\wideparen{k}}+\cdots + P_i P_j\cdots P_{m} \vec{C}_{n|\wideparen{n}}\,.
\label{conc_relation}
\eea
This shows that the information contained in the $N$ elementary concurrence vectors is enough to built any other concurrence vector, and thus the corresponding concurrence, ${C}_{I|\wideparen{I}}^2=\|\vec{C}_{I|\wideparen{I}}\|^2$.

\section{Triangular inequalities}\label{sec: triangular}

Consider a tripartition of the Hilbert space $\{i,j,\wideparen{ij}\}$. The corresponding concurrence vectors are related by 
Eq.(\ref{2to1}), which can be written as
\be
P_j\vec{C}_{ij|\wideparen{ij}}=\vec{C}_{i|\wideparen{i}}-\vec{C}_{j|\wideparen{j}}\,.
\label{simplecase}
\ee
As these three vectors form a triangle, their lengths satisfy the triangular inequality. Since $\|P_i\vec{C}_{ij|\wideparen{ij}}\|=\|\vec{C}_{ij|\wideparen{ij}}\|=C_{ij|\wideparen{ij}}$, the inequality reads
\be
C_{ij|\wideparen{ij}}\leq 
C_{i|\wideparen{i}}+C_{j|\wideparen{j}}
\label{triangular}
\ee
(and permutations of the three terms). This triangular inequality has been shown in previous literature, e.g. in refs \cite{ PhysRevA.102.062424, Jin_2023}. With the use of concurrence vectors it is a quite trivial result since
the involved vectors form a triangle (with a permutation of components in one of them). This provides a  transparent geometrical interpretation of the triangular inequality.

We can also show very easily the (stronger) inequality for the squared of the concurrences \cite{PhysRevA.92.062345},
\be
C_{ij|\wideparen{ij}}^2\leq 
C_{i|\wideparen{i}}^2+C_{j|\wideparen{j}}^2
\label{pitagoras}
\ee
(and permutations). To see this notice that
\bea
C_{ij|\wideparen{ij}}^2=\|(\mathds{1}-P_iP_j)\vec{A}\|^2=
\|(\mathds{1}-P_i)\vec{A}-(\mathds{1}-P_j)\vec{A}\|^2=
C_{i|\wideparen{i}}^2+C_{j|\wideparen{j}}^2 -2\vec{A}^\dagger
(\mathds{1}-P_i)(\mathds{1}-P_j)\vec{A}\,,
\label{a1}
\eea
where the last term is negative semidefinite:
\be
-2\vec{A}^\dagger
(\mathds{1}-P_i)(\mathds{1}-P_j)\vec{A}=
-\frac{1}{2}\vec{A}^\dagger
\big((\mathds{1}-P_i)(\mathds{1}-P_j)\big)^2\vec{A}=-\frac{1}{2}\|(\mathds{1}-P_i)(\mathds{1}-P_j)\vec{A}\|^2\leq 0\,.
\label{a2}
\ee

\vspace{0.2cm}
Clearly, relations (\ref{triangular}, \ref{pitagoras}) hold for any tripartition of the Hilbert space, i.e. the indices $i, j$ can be replaced by collective indices $I,J$ when $I\cap J=\emptyset$. 
If $I\cap J\neq\emptyset$ we simply note that
\bea
\vec{C}_{I\vartriangle J|\wideparen{I\vartriangle J}}=(\mathds{1}-P_IP_J)\vec{A}= P_J\Big((\mathds{1}-P_I)-(\mathds{1}-P_J)\Big)\vec{A} = 
P_J\left(\vec{C}_{I|\wideparen{I}}-\vec{C}_{J|\wideparen{J}}\right)\,,
\eea
where $I\vartriangle J=I\cup J \setminus I\cap J$ is the set of indices in either $I$ or $J$, but not in both. Hence, analogous expressions to Eqs.(\ref{a1}, \ref{a2}) can be obtained,
\begin{eqnarray}\nn
    &\dis{C_{I\vartriangle J|\wideparen{I\vartriangle J}}^2=
    C_{I|\wideparen{I}}^2+C_{J|\wideparen{J}}^2 -2\vec{A}^\dagger
    (\mathds{1}-P_I)(\mathds{1}-P_J)\vec{A}
    }\,, \\ \ecart 
    &\dis{-2\vec{A}^\dagger
    (\mathds{1}-P_I)(\mathds{1}-P_J)\vec{A}=
    -\frac{1}{2}\|(\mathds{1}-P_I)(\mathds{1}-P_J)\vec{A}\|^2\leq 0}\,,
    \label{aGen}
\end{eqnarray}
and in consequence the triangular inequalities (\ref{triangular}, \ref{pitagoras}) generalize to non-disjoint subsystems as
\bea
C_{I\vartriangle J|\wideparen{I\vartriangle J}}&\leq& 
C_{I|\wideparen{I}}+C_{J|\wideparen{J}}\,,
\nonumber\\
C_{I\vartriangle J|\wideparen{I\vartriangle J}}^2&\leq& 
C_{I|\wideparen{I}}^2+C_{J|\wideparen{J}}^2\,.
\label{general_triangular}
\eea

On the other hand, in ref. \cite{Xie_2021} it was shown that, {\em if} the system consists of three qubits, the inequality (\ref{pitagoras}) becomes an equality only when either $C_{i|\wideparen{i}}$ or $C_{j|\wideparen{j}}$ are vanishing; otherwise the $``\leq"$ sign in (\ref{pitagoras}) becomes strict inequality $``<"$.
In Appendix A we show that this interesting result holds for any tripartition, independently of the dimension of the subsystems. As a consequence, the area of the ``concurrence triangle" built up with the squared concurrences of a tripartite system is a consistent measure of genuine entanglement. This entanglement measure was advocated in ref.\cite{Xie_2021,Jin_2023} for systems of three qubits, but this result shows that it can be extended to arbitrary dimension of the subsystems. Besides, the same property holds for non-disjoint tripartitions, i.e. Eq.(\ref{general_triangular}) becomes an strict equality {\em iff} either 
$C_{I|\wideparen{I}}$ or $C_{J|\wideparen{J}}$ are vanishing.

Finally, using the general expression (\ref{conc_relation}) one can go beyond tripartitions and derive other inequalities. An obvious one comes from the fact that the vectors involved in (\ref{conc_relation}) form a (non-necessarily flat) polygon, so they satisfy the polygonal inequality (i.e. the extension of the triangular inequality to polygons). Since the permutations do not modify the norm of the various vectors involved, we conclude that 
\bea
{C}_{I|\wideparen{I}} \leq
{C}_{i|\wideparen{i}}+{C}_{j|\wideparen{j}}+\cdots + {C}_{m|\wideparen{m}}\,,
\nonumber\\
{C}_{I|\wideparen{I}}^2 \leq
{C}_{i|\wideparen{i}}^2+{C}_{j|\wideparen{j}}^2+\cdots + {C}_{m|\wideparen{m}}^2\,,
\eea
as well as
\bea
{C}_{I\vartriangle J\vartriangle \cdots \vartriangle M|\wideparen{I\vartriangle J\vartriangle \cdots \vartriangle M}} \leq
{C}_{I|\wideparen{I}}+{C}_{J|\wideparen{J}}+\cdots + {C}_{M|\wideparen{M}}\,,
\nonumber\\
{C}_{I\vartriangle J\vartriangle \cdots \vartriangle M|\wideparen{I\vartriangle J\vartriangle \cdots \vartriangle M}}^2 \leq
{C}_{I|\wideparen{I}}^2+{C}_{J|\wideparen{J}}^2+\cdots + {C}_{M|\wideparen{M}}^2
\eea
(and permutations). 

A straightforward way to get new inequalities among concurrences is to consider the inequality
\bea
\vec{A}^\dagger
(\mathds{1}-P_I)(\mathds{1} \pm P_J)(\mathds{1} \pm P_K)\cdots\ (\mathds{1} \pm P_M)\vec{A}\geq 0\,.
\label{gen_ineq}
\eea
Since $(\mathds{1} \pm P)=(\mathds{1} \pm P)^2/2$, the l.h.s. of this equation is positive semidefinite, and it can be expressed as a combination of concurrences.

\section{Subadditivity and strong subadditivity}\label{sec:entropy}
 
From an entropy-of-entanglement perspective, the concurrence can be identified (up to a normalization factor) with the so-called Tsallis-$2$ (or ``linear") entropy \cite{Tsallis:1987eu, 10.1063/1.2165794, LANDSBERG1998211}
\be
C^2_{I|\wideparen{I}}=2
\left(1-\tr\rho_{A}^2\right) \equiv 2\ S_2\left(\rho_{A}\right)\,,
\label{Tsallis}
\ee
where we have kept the notation previously introduced. Some of the well-known properties of 
the von Neumann  entropy, $S_{\rm VN}(\rho)=-\tr\left(\rho \log \rho\right)$, are also shared  by $S_2(\rho)$\cite{Rastegin2011}. In particular, given two subsystems, ${\cal H}_{ A}\otimes {\cal H}_{ B}$,
the $S_2-$entropy satisfies the {\em subadditivity} conditions
\cite{10.1063/1.530920, Audenaert2007}:
\be
|S_2\left(\rho_{A}\right)-S_2\left(\rho_{B}\right)|\leq S_2\left(\rho_{AB}\right)\leq S_2\left(\rho_{A}\right)+S_2\left(\rho_{B}\right)\,.
\label{subadditivity}
\ee
The previous second inequality actually corresponds to the positivity of an information-theory quantity known as mutual information, 
associated with the total amount of correlations in $\rho_{AB}$
\cite{Groisman_2005}
\begin{equation}
    I\left(A: B\right) \equiv S_2\left(\rho_{A}\right)+S_2\left(\rho_{B}\right)-S_2\left(\rho_{AB}\right)\geq0\,.
    \label{info}
\end{equation}

Now, these results stem trivially from the triangular inequalities (\ref{triangular}, \ref{pitagoras}) by simply considering a tripartition of the Hilbert space, ${\cal H} = {\cal H}_{ A}\otimes {\cal H}_{ B}\otimes {\cal H}_{\wideparen{AB}}$, and identifying the indices $i,j$ with the subsystems ${\cal H}_{ A}\otimes {\cal H}_{ B}$. We stress that the subadditivity condition holds for an arbitrary density matrix $\rho_{AB}$. In this context, ${\cal H}_{\wideparen{AB}}$ can be considered as an extension of the Hilbert space ${\cal H}_{ A}\otimes {\cal H}_{ B}$, so that the global state becomes pure, which is always possible thanks to the purification theorem. 

In addition, we observe the following interesting property. Since the triangular inequality (\ref{pitagoras}) becomes an equality {\em iff} either $C_{i|\wideparen{i}}$ or $C_{j|\wideparen{j}}$ are vanishing (see discussion in section \ref{sec: triangular} and a proof in Appendix A), it follows that the subadditivity condition (\ref{info}) is saturated
{\em iff} the entropy of one of the two subsystems is vanishing:
\begin{equation}
S_2\left(\rho_{A}\right)+S_2\left(\rho_{B}\right)-S_2\left(\rho_{AB}\right)=0\ \iff\ S_2\left(\rho_{A}\right)=0\ \ {\rm or}\ \  S_2\left(\rho_{B}\right)=0\,.
    \label{subadequality}
\end{equation}

Further relations concerning the Tsallis-2 entropy have been tested and, among them, special attention has been paid to the {\em strong} subadditivity \cite{Lieb2018}. For a generic entropy, $S(\rho)$, strong subadditivity reads
\bea
S\left(\rho_{ABC}\right)+ S\left(\rho_{B}\right)\leq S\left(\rho_{AB}\right)+S\left(\rho_{BC}\right),\ \  {\rm i.e.}\ \   I\left(A : B\right)\leq I\left(A : BC\right)\,,
\label{strong1}
\eea
where the last expression indicates that correlations are non-decreasing under the extension of one of the subsystems. 
It is well known that von Neumann entropy, $S_{\rm VN}(\rho)$, satisfies the strong subadditivity condition \cite{Lieb2018}.
In contrast, generically the Tsallis-2 entropy, $S_2(\rho)$, does {\em not} satisfy it, as it was proven in ref. \cite{Petz:2015gxk} by constructing several counterexamples. We can easily check this result using the concurrence-vector approach. Consider a Hilbert space partitioned as ${\cal H} = {\cal H}_{ A}\otimes {\cal H}_{ B}\otimes{\cal H}_{ C}\otimes {\cal H}_{\wideparen{ABC}}$, i.e. $\{i,j,k,\wideparen{ijk}\}$. 
In terms of concurrences, condition (\ref{strong1}) for $S_2(\rho)$ {\em would} read
\bea
C_{ijk|\wideparen{ijk}}^2+C_{j|\wideparen{j}}^2-
C_{ij|\wideparen{ij}}^2-C_{jk|\wideparen{jk}}^2\ \leq \ 0\,.
\label{strong2}
\eea
However, the l.h.s. of this relation is exactly the quantity: 
\bea
&&2\vec{A}^\dagger\left((\mathds{1}-P_iP_jP_k)+(\mathds{1}-P_j)-(\mathds{1}-P_iP_j)-(\mathds{1}-P_jP_k)\right)\vec{A}\phantom{\frac{1}{1}}
\nonumber\\
&&=\ -2\vec{A}^\dagger P_j (\mathds{1}-P_i)(\mathds{1}-P_k)\vec{A} \phantom{\frac{1}{1}}\,,
\label{viol}
\eea
which is not negative semidefinite, showing that strong subadditivity (\ref{strong2}) can be violated for particular choices of the state $\vec{A}$. E.g. for  $(\mathds{1}-P_i P_j)\vec{A}=0$, i.e. $C_{ij|\wideparen{ij}}^2=0$, the expression (\ref{viol}) becomes
\begin{equation}
2\vec{A}^\dagger
  (\mathds{1}-P_j)(\mathds{1}-P_k) \vec{A}\ \geq\ 0\,.
\end{equation}
This inequality is strict {\em iff}, moreover, $C_{j|\wideparen{j}}^2\ $, $C_{k|\wideparen{k}}^2\neq0$ (see discussion after Eq.(\ref{general_triangular}) and Appendix A), thus leading to the violation of the strong subadditivity.
An example of this setup is a state $|\text{Bell}\rangle_{AB}\otimes |\text{Bell}\rangle_{C\wideparen{ABC}}\in
{\cal H}_{ A}\otimes {\cal H}_{ B}\otimes{\cal H}_{ C}\otimes {\cal H}_{\wideparen{ABC}}$.

On the other hand, the quantity (\ref{viol}) becomes obviously negative semidefinite for $P_j \vec{A} = \vec{A}$, i.e. $C_{j|\wideparen{j}}^2=0$,
\be
-2\vec{A}^\dagger (\mathds{1}-P_i)(\mathds{1}-P_k)\vec{A} \leq 0\,,
\ee
so in this case the strong subadditivity is fulfilled.
Actually, this instance is equivalent to the
``ordinary" subadditivity (\ref{subadditivity}) for the tripartition ${\cal H} = {\cal H}_{ A}\otimes {\cal H}_{ C}\otimes {\cal H}_{\wideparen{AC}}$.

Notice also that for $C_{i|\wideparen{i}}^2=0$ or $C_{k|\wideparen{k}}^2=0$ Eq.(\ref{viol}) vanishes, so the strong subadditivity holds as an equality.

\vspace{0.2cm}
To finish this section, let us
construct a modified version of the strong subadditivity which does always hold for  the Tsallis-2 entropy (\ref{Tsallis}). Expanding the inequality (see Eq.(\ref{gen_ineq}))
\be
\vec{A}^\dagger(\mathds{1}-P_i)(\mathds{1}+P_j)(\mathds{1}-P_k)\vec{A}\geq 0
\ee
we get
\bea
C_{ijk|\wideparen{ijk}}^2+C_{j|\wideparen{j}}^2 \leq
C_{ij|\wideparen{ij}}^2+C_{jk|\wideparen{jk}}^2+\left(C_{i|\wideparen{i}}^2+C_{k|\wideparen{k}}^2 -C_{ik|\wideparen{ik}}^2\right)\,,
\eea
where the term within brackets is positive semidefinite, see Eq.(\ref{pitagoras}), and represents the departure from the  strong-subadditivity condition (\ref{strong2}). 
In terms of the Tsallis-2 entropy this softened version of strong subadditivity reads
\bea\nn
S_2\left(\rho_{ABC}\right)+ S_2\left(\rho_{B}\right)&\leq &S_2\left(\rho_{AB}\right)+S_2\left(\rho_{BC}\right)
+\Big[S_2\left(\rho_{A}\right)+S_2\left(\rho_{C}\right)-S_2\left(\rho_{AC}\right)\Big]\,,
\label{strong_modified}
\eea
where the positivity (semi)definiteness of
the term in brackets is equivalent to the ordinary subadditivity condition (\ref{subadditivity}). 
In terms of mutual information this relation reads 
$I\left(A : B\right)\leq I\left(A : BC\right)+I\left(A : C\right)$. Since we can exchange $B\leftrightarrow C$, it can be expressed as
\be
\left| I\left(A : B\right)-I\left(A : C\right)\right|
\leq I\left(A : BC\right)\,.
\ee

On the other hand, we can yet construct an alternative subadditivity using the general relation (\ref{general_triangular}). By considering a Hilbert space partitioned as ${\cal H} = {\cal H}_{ A}\otimes {\cal H}_{ B}\otimes{\cal H}_{ C}\otimes {\cal H}_{\wideparen{ABC}}$,
and identifying $I=\{i,j\}$, $J=\{j,k\}$, we get
\be
C_{ik|\wideparen{ik}}^2\leq C_{ij|\wideparen{ij}}^2 + C_{jk|\wideparen{jk}}^2\,.
\label{triangular222}
\ee
In terms of entropies:
\bea
S_2\left(\rho_{AC}\right)\leq S_2\left(\rho_{AB}\right)+S_2\left(\rho_{BC}\right)\,,
\eea
which is a kind of triangular inequality for the $S_2-$entropies. Analogously to the ordinary subadditivity, the fact that relation (\ref{triangular222}) 
becomes an equality {\em iff} either $C_{ij|\wideparen{ij}}$ or $C_{jk|\wideparen{jk}}$ are vanishing (see Appendix A), it follows that 
\bea
S_2\left(\rho_{AC}\right)= S_2\left(\rho_{AB}\right)+S_2\left(\rho_{BC}\right)\ \iff\ S_2\left(\rho_{AB}\right)=0\ \ {\rm or}\ \  S_2\left(\rho_{BC}\right)=0\,.
\eea
Concerning the mutual information, further inequalities can also be derived. In particular, applying \eq{gen_ineq} to the combination $(\mathds{1}-P_i)(\mathds{1}-P_j)(\mathds{1}-P_k)$ we obtain a non-negative condition over the tripartite information \cite{Casini:2008wt}
\begin{eqnarray}\nn
    S_2\left(\rho_{AB}\right)+S_2\left(\rho_{BC}\right)+S_2\left(\rho_{AC}\right)&\leq& S_2\left(\rho_{A}\right)+ S_2\left(\rho_{B}\right)+ S_2\left(\rho_{C}\right)+ S_2\left(\rho_{ABC}\right) \\ \ecart
    \Rightarrow\  
    I\left(A : B: C\right)&\geq& 0\,,
\end{eqnarray}
where $I\left(A : B: C\right)
=I\left(A : B\right)+I\left(A : C\right)-I\left(A : BC\right)$. The sign of this quantity for the von Neumann entropy has been explored in the literature \cite{Casini:2008wt}. For the case of holographic theories it has been proven to be non-positive (Monogamy of Mutual Information) \cite{Hayden:2011ag}, i.e. exactly the opposite behavior as the one found above for the Tsallis-2 entropy. The same relation in a similar context was obtained in refs.\cite{Eltschka_2018a, Eltschka_2018b}.

\section{Sufficient conditions for genuine entanglement}\label{sec: suf_conds}

Let us now use the idea of concurrence vector and its expression (\ref{convec4}) to derive new sufficient conditions for genuine entanglement, which are computable in polynomial time.

For the sake of the clarity of the discussion, let us denote $P_1, P_2, \dots P_N$ the elementary permutations associated with the indices $i, j, k, \dots$, corresponding to the Hilbert spaces ${\cal H}_1\otimes {\cal H}_2\otimes \cdots \otimes {\cal H}_N$.

\vspace{0.3cm}
\noindent
\underline{\bf $N$ even}

\vspace{0.2cm}
Consider the vector
\be
\vec{V}_N=(\mathds{1}-P_1)(\mathds{1}-P_2)\cdots (\mathds{1}-P_{N-1}) \vec{A}\,,
\label{VN}
\ee
which is a linear combination of all the concurrences:
\be
\vec{V}_N=\sum_{I\subset \{1,2, \dots N-1\}} (-1)^{\#I} (\mathds{1}-P_I)\vec{A}\ \ =\sum_{I\subset \{1,2, \dots N-1\}} (-1)^{\#I} \vec{C}_{I|\wideparen{I}}\,,
\label{VN2}
\ee
where $\#I$ is the cardinality of $I$.
The subscript $N$ indicates that the $(\mathds{1}-P_N)$ factor has been left outside the product in Eq.(\ref{VN}).
In this way we avoid the duplication of concurrences, since $(\mathds{1}-P_I)\vec{A}$ and $(\mathds{1} -
P_{\wideparen{I}})
\vec{A}$ are equal and correspond to the same bipartition.
Now suppose that there exists a permutation $P_\Sigma=P_{\sigma_1}P_{\sigma_2}\cdots P_{\sigma_m}$ with $m$ odd, such that
\be
(\mathds{1}-P_\Sigma)\vec{A}=0\,.
\ee
In other words, suppose that the state is separable along the bipartition $\Sigma|\wideparen{\Sigma}$. Without loss of generality, we can suppose $\sigma_1,\sigma_2, \cdots \sigma_m \leq N-1$ (otherwise, consider the equivalent permutation $P_{\wideparen{\Sigma}}$ instead of $P_\Sigma$). Then, Eq.(\ref{VN}) vanishes since the terms of the sum can be paired as:
\be
\vec{V}_N=\frac{1}{2}\sum_{I\subset \{1,2, \dots N-1\}} (-1)^{\#I} \left((\mathds{1}-P_I) - (\mathds{1}-P_IP_\Sigma)\right)\vec{A}=0\,.
\label{VN=0}
\ee
In consequence, if $\vec{V}_N\neq 0$, this is a sufficient condition to exclude any separable ``odd" bipartition. It is remarkable that this check of the $\sim 2^{N-2}$ odd bipartitions is carried out by 
$N-1$ simple operations on $\vec{A}$.

The previous condition though says nothing about the possibility of $(\mathds{1}-P_\Sigma)\vec{A}=0$ where $P_\Sigma=P_{\sigma_1}P_{\sigma_2}\cdots P_{\sigma_m}$ with $m$ even (again we can assume $\sigma_1,\sigma_2, \cdots \sigma_m \leq N-1$).
To probe this instance we construct the following $N-1$ vectors
\bea
\vec{W}^{(1)}_N&=&(\mathds{1}+P_1)(\mathds{1}-P_2)\cdots (\mathds{1}-P_{N-1}) \vec{A}\,,
\nonumber\\
\vec{W}^{(2)}_N&=&(\mathds{1}-P_1)(\mathds{1}+P_2)\cdots (\mathds{1}-P_{N-1}) \vec{A}\,,
\nonumber\\
&\vdots&
\nonumber\\
\vec{W}^{(N-1)}_N&=&(\mathds{1}-P_1)(\mathds{1}-P_2)\cdots (\mathds{1}+P_{N-1}) \vec{A}\,.
\label{WN}
\eea
I.e., $\vec{W}^{(k)}_N$ is like $\vec{V}_N$, with the sign of $P_k$ flipped. It can also be expressed as a linear combination of concurrences:
\be
\vec{W}^{(k)}_N=\sum_{I\subset \{1,2, \dots N-1\}} (-1)^{[I]_k}(-1)^{\#I} (\mathds{1}-P_I)\vec{A}\ \ =\sum_{I\subset \{1,2, \dots N-1\}} (-1)^{[I]_k}(-1)^{\#I} \vec{C}_{I|\wideparen{I}}\,,
\ee
where $[I]_k=1$ ($0$) for $k\in I$ ($k\notin I$).
Now, {\em if} $k\in\Sigma$, the previous expression can be cast as 
\be
\vec{W}^{(k)}_N=\frac{1}{2}\sum_{I\subset \{1,2, \dots N-1\}} (-1)^{[I]_k}(-1)^{\#I} \big[(\mathds{1}-P_I)-(\mathds{1}-P_IP_\Sigma)\big]\vec{A}\ =\ 0\,.
\ee
Consequently, if $(\mathds{1}-P_{\sigma_1}P_{\sigma_2}\cdots P_{\sigma_m})\vec{A}=0$, with $m$ even, at least one $\vec{W}^{(k)}_N$ must vanish. Correspondingly, if all $\vec{W}^{(k)}_N\neq 0$, this suffices to exclude the possibility of a separable even bipartition. Note that the condition for $m$ even requires $(N-1)^2$ operations, still a polynomial task.

In summary, the conditions 
$\vec{V}_N\neq 0$, 
$\vec{W}^{(k)}_N\neq 0\  \forall k$ are sufficient to guarantee that the state is genuinely entangled.

\vspace{0.3cm}
\noindent
\underline{\bf $N$ odd}

\vspace{0.2cm}
Suppose that there exists a permutation $P_\Sigma=P_{\sigma_1}P_{\sigma_2}\cdots P_{\sigma_m}$ such that
\be
(\mathds{1}-P_\Sigma)\vec{A}=0\,.
\ee
Without loss of generality, we can suppose $m$ odd
(otherwise, consider the equivalent permutation $P_{\wideparen{\Sigma}}$ instead of $P_\Sigma$). Now, if 
$\sigma_1,\sigma_2, \cdots \sigma_m \leq N-1$, then $\vec{V}_N$, defined by Eq.(\ref{VN}), vanishes, as discussed around Eq.(\ref{VN=0}). Of course, it may happen that not all $\sigma_1,\sigma_2, \cdots \sigma_m \leq N-1$. However, if we consider the following $N$ vectors
\bea
\vec{V}_1&=&(\mathds{1}-P_2)(\mathds{1}-P_3)\cdots (\mathds{1}-P_{N}) \vec{A}\,,
\nonumber\\
\vec{V}_2&=&(\mathds{1}-P_1)(\mathds{1}-P_3)\cdots (\mathds{1}-P_{N}) \vec{A}\,,
\nonumber\\
&\vdots&
\nonumber\\
\vec{V}_N&=&(\mathds{1}-P_1)(\mathds{1}-P_2)\cdots (\mathds{1}-P_{N-1}) \vec{A}\,,
\label{VN_gen}
\eea
it is clear that all the indices $\sigma_1,\sigma_2, \cdots \sigma_m$ will be contained in at least one of them. Then, at least one $\vec{V}_k=0$.

Consequently if all $\vec{V}_k\neq 0$, then the state is genuinely entangled. This check accomplishes $N^2$ operations.

Incidentally, for $N=3$ the previous conditions, i.e. $\vec{V}_1\neq 0$, $\vec{V}_2\neq 0$, are also necessary conditions for genuine entanglement since, as discussed in section \ref{sec: triangular}, $(\mathds{1}-P_i)(\mathds{1}-P_j)\vec{A}=0 \iff (\mathds{1}-P_i)\vec{A}=0 \ {\rm or}\  (\mathds{1}-P_j)\vec{A}=0$.

\section{Summary and conclusions}\label{conclusions}

Quantum entanglement and quantum entropy are crucial concepts in the study of multipartite quantum systems. In this work we have shown how the notion of concurrence vector, Eq.(\ref{convec}),  re-expressed in the particular useful form (\ref{convec4}) (which is central for the present work), provides new insights and computational tools for the analysis of both. 

In a multipartite system, states with genuine entanglement are by definition those that are entangled under any possible bipartition. Using the approach described here
for a general multipartite pure state, it is easy to obtain equalities that relate the concurrence vectors of different bipartitions, and thus about the genuine entanglement of the system. On the one hand, this provides interesting geometrical insights, e.g. the well-known triangular inequality for concurrences arises from the trivial fact that the corresponding concurrence vectors form a triangle. On the other hand, it allows to prove known relations in an easy way and to build up new ones. An example of the latter is the fact that the triangular inequality $C_{AB|\wideparen{AB}}^2\leq C_{A|\wideparen{A}}^2+C_{B|\wideparen{B}}^2$ becomes an equality {\em iff} $C_{A|\wideparen{A}}=0$ or $C_{B|\wideparen{B}}=0\ $; a result that was proven in ref.\cite{Xie_2021} only for the three-qubit case. In consequence, the area of the triangle built up with the squared concurrences is a sound measurement of genuine entanglement for generic tripartite systems, as was proposed in refs.\cite{Xie_2021,Jin_2023} for the three-qubit case.
The present concurrence-vector approach is also useful to derive sufficient conditions for entanglement in generic multipartite systems that are computable in polynomial time.

From an entropy-of-entanglement perspective, the concurrence of a bipartite system in a pure state can be identified (up to a normalization factor) with the so-called Tsallis-$2$ entropy, $S_2(\rho_A)$, in one of the two subsystems. This allows to use the present approach to easily prove relations such as the subadditivity of $S_2$. Likewise, it is a useful tool to prove new relations, e.g. a modified version of the strong subadditivity which is always fulfilled. Thanks to the purification theorem all these results hold for any multipartite state, whether pure or mixed.

All this shows that the concurrence vector approach is a useful tool for the study of the entanglement and the entropy of multipartite systems.

\section*{Acknowledgements}

	We are grateful to  G. Sierra for useful discussions. The authors acknowledge the support of the Spanish Agencia Estatal de Investigacion through the grants ``IFT Centro de Excelencia Severo Ochoa CEX2020-001007-S" and PID2019-110058GB-C22 funded by MCIN/AEI/10.13039/501100011033 and by ERDF. The work of A.B. is supported through the FPI grant PRE2020-095867 funded by MCIN/AEI/10.13039/501100011033.

\subsubsection*{Data availability}
All data and material considered in this paper are included in the text.

\subsubsection*{Conflict of interest}
The authors declare that they do not have conflict of interest.

\newpage

\section{Appendix A. The equality in the triangular relation}

As mentioned in section \ref{sec: triangular}, for a system of {\em three qubits} the triangular inequality
\be
C_{ij|\wideparen{ij}}^2
\leq 
C_{i|\wideparen{i}}^2+C_{j|\wideparen{j}}^2
\label{A1}
\ee
becomes an equality {\em iff} either $C_{i|\wideparen{i}}=0$ or $C_{j|\wideparen{j}}=0$. This interesting result was proven in ref.\cite{Xie_2021}. Here we extend it to any tripartite system or any tripartition, independently of the dimensions of the involved Hilbert spaces. 

From Eqs.(\ref{a1}, \ref{a2}) the equality in the relation (\ref{A1}) is equivalent to the condition
\be
(\mathds{1}-P_i)(\mathds{1}-P_j)\vec{A}=0\,,
\label{A3}
\ee
where $\vec{A}$ is the vector of the coefficients of the state $|\psi\rangle \otimes |\psi\rangle \in {\cal H}\otimes{\cal H}$, see Eq.~(\ref{def_A}). Taking into account that for any bipartition $I|\wideparen{I}$ the concurrence reads $C_{I|\wideparen{I}}^2=\|(\mathds{1}-P_i)\vec{A}\|^2$, our goal is to prove that Eq.(\ref{A3}) requires $(\mathds{1}-P_i)\vec{A}=0$ or $(\mathds{1}-P_j)\vec{A}=0$. 

\subsection{3 qubits and 3 qutrits}

Let us first consider the case of three qubits, so that
\be
A_{i j k ;\, i' j' k'}=(a_{i j k})\, (a_{i' j' k'})
\label{vecA}
\ee
with all indices taking two possible values, $0$ and $1$. Thus, in this case Eq.(\ref{A3}) represents $(2^3)^2$ quadratic equations in the $a_{ijk}$ variables. Actually, most of them are trivial, i.e. the l.h.s. of (\ref{A3}) is identically zero. For the non-trivial ones has always  the form $\pm q_0,\  \pm q_1$ or $\pm q_2$, with
\begin{eqnarray}
q_0 &=&\Bopo \Bpoo- \Booo \Bppo\,,  \nonumber  \\
q_1 &= &\Bopp \Bpop- \Boop \Bppp\,, \\ 
q_2 &=&\Bopp \Bpoo+\Bopo \Bpop-\Boop \Bppo-\Booo \Bppp \nonumber
\end{eqnarray}
(notice that  $q_1 =   q_0 \displaystyle|_{\Bijo \longleftrightarrow \Bijp} $). Consequently, Eq.(\ref{A3}) is equivalent to 
\be
\{ q_0, q_1, q_2\}=\{0,0,0\}\,.
\label{qs}
\ee
Incidentally, a similar result was obtained in refs.~\cite{Uskov_2008, Uskov_2020} for different linear combinations of these vectors.
By a careful (and lengthy) inspection it is possible to check that the only consistent possibility to fulfill Eq.(\ref{qs}) is indeed that either $(\mathds{1}-P_i)\vec{A}=0$ or $(\mathds{1}-P_j)\vec{A}=0$. However, a more expeditious way to show this is the following. Using {\sf Singular}, a computer algebra system for polynomial computations \cite{DGPS} it is easy to prove that 
\begin{equation}
\|(\mathds{1}-P_1)\vec{A}\|^2    \|(\mathds{1}-P_2) \vec{A}\|^2 =  s_0 q_0 +  s_1 q_1  +   s_2 q_2 
\end{equation}
for some -sextic- polynomials  $\{ s_0, s_1, s_2\}$ on the variables $a_{ijk}, a_{ijk}^*$. (The explicit form of $s_0,s_1,s_2$ is quite longish and of no particular interest, so we omit it.). Thus, indeed, Eq.(\ref{A3}) requires $(\mathds{1}-P_i)\vec{A}=0$ or $(\mathds{1}-P_j)\vec{A}=0$ and the statement is proven for three qubits. 

A similar computation shows that the same statement holds for a system of three qutrits, i.e. when the indices of $\vec{A}$ in Eq.(\ref{vecA}) take the values $0,1,2$. In this case, the calculation is more involved: Eq.(\ref{A3}) represents $(3^3)^2$ quadratic equations, 54 of which are non-equivalent. The complexity of Eq.(\ref{A3}) grows geometrically with the dimension of the Hilbert spaces of the three subsystems. However, the result for three qutrits is all we need to recursively extend statement to any dimension, as shown below. 

\subsection{Extension to arbitrary dimension}

The above result for three qutrits can be re-stated in the following way. Given, by hypothesis, Eq. (\ref{A3}), if $(\mathds{1}-P_j)\vec{A}\neq 0$ for some component(s) of $\vec{A}$, then necessarily $(\mathds{1}-P_i)\vec{A}=0$ for all components. (Of course the same holds for $i\leftrightarrow j$.)

It is convenient at this point to define the restriction of the vector $\vec{A}$ to a subset of components, i.e. to the subspace spanned by a subset of indices. E.g.
\be
   \vec{A}^{(i_1 i_2)\,(j_1 j_2)\,(k_1 k_2)} = \{a_{i j k}\,a_{i' j' k'}\}\ \ \ {\rm with}\ \ \ i,i'\in\{i_1, i_2\},\, j,j'\in\{j_1, j_2\},\, k,k'\in\{k_1, k_2\}\,.
 \ee
Note that the action of $P_i, P_j$ or $P_k$ on this restricted vector keeps it inside the same subspace. In this notation, the previous result for qutrits reads as follows. Given a state such that $(\mathds{1}-P_i)(\mathds{1}-P_j)\vec{A}= 0$ and 
\be
(\mathds{1}-P_j)\vec{A}^{(i_1 i_2)\,(j_1 j_2)\,(k_1 k_2)}\neq 0
\label{stqub}
\ee
(for some component(s)), then
\be
(\mathds{1}-P_i)\vec{A}^{(i_1 i_2 i_3)\,(j_1 j_2 j_3)\,(k_1 k_2 k_3)}= 0\,.
\label{stqut}
\ee
To extend this result to higher dimensions, let us consider the multipartite Hilbert space
\be
{\cal H} = {\cal H}_{1}\otimes {\cal H}_{2}\otimes{\cal H}_{3}\,,
\ee
with ${\rm dim}\, {\cal H}_n=d_n\geq 4$. Our hypothesis is again Eq.(\ref{A3}) and Eq.(\ref{stqub}) in some subspace corresponding to the indices $\{(i_1 i_2)\,(j_1 j_2)\,(k_1 k_2)\}$, i.e. some subspace of three qubits.
Let us consider the addition of a fourth set of indices $\{i_4 j_4 k_4\}$.
Our goal is to show that
\be
(\mathds{1}-P_i)\vec{A}^{(i_1 i_2 i_3 i_4)\,(j_1 j_2 j_3 j_4)\,(k_1 k_2 k_3 k_4)}= 0\,.
\label{stfour}
\ee
Since we have shown that we can go from (\ref{stqub}) to Eq.(\ref{stqut}) for {\em any} new indices $i_3, j_3, k_3$, it follows that Eq.(\ref{stfour}) is satisfied by many subsets of components, namely those corresponding to extend the three qubits of Eq.(\ref{stqub}) to any three qutrits:
\bea
(\mathds{1}-P_i)\vec{A}^{(i_1 i_2 i_3)\,(j_1 j_2 j_3)\,(k_1 k_2 k_3)}&=& 0\,,
\nonumber\\
(\mathds{1}-P_i)\vec{A}^{(i_1 i_2 i_3)\,(j_1 j_2 j_3 )\,(k_1 k_2 k_4)}&=& 0\,,
\nonumber\\
(\mathds{1}-P_i)\vec{A}^{(i_1 i_2 i_3)\,(j_1 j_2 j_4)\,(k_1 k_2 k_4)}&=& 0\,,
\label{various}
\\
{\rm etc.}\hspace{3cm} &&
\nonumber
\eea
Hence, the only components to check are those $A_{i j k ;\, i' j' k'}=(a_{i j k})\, (a_{i' j' k'})$ where $i=i_3, i'=i_4$ (or vice versa) and/or $j=j_3, j'=j_4$ (or vice versa) and/or $k=k_3, k'=k_4$ (or vice versa). Let $A_{pqr; p'q'r'}=
(a_{p q r})\, (a_{p' q' r'})$ be one of such components. We have to show that $(\mathds{1}-P_i)A_{pqr; p'q'r'}=0$, i.e.
\be
(a_{p q r})\, (a_{p' q' r'})=(a_{p' q r})\, (a_{p q' r'})\,.
\ee
From Eq.(\ref{stqub}), there must be some component of $\vec{A}^{(i_1 i_2)\,(j_1 j_2)\,(k_1 k_2)}$ different from zero. Suppose $A_{i_1 j_1 k_1 ;\, i_2 j_2 k_2}=(a_{i_1 j_1 k_1})\, (a_{i_2 j_2 k_2})\neq 0$ (the argument goes the same for any other component). Then the following chain of equalities follow from Eq.(\ref{various}):
\begin{eqnarray}
    &(a_{i_1 j_1 k_1})\, (a_{i_2 j_2 k_2})\, (a_{p q r})\, (a_{p' q' r'})& \\ \ecart \nn
    =&(a_{p j_1 k_1})\, (a_{p' j_2 k_2})\, (a_{i_1 q r })\, (a_{i_2 p' r'})& \\ \ecart \nn
    =&(a_{i_2 j_1 k_1})\, (a_{i_1 j_2 k_2})\, (a_{p' q r})\, (a_{p q' r' })& \\ \ecart \nn
    =&(a_{i_1 j_1 k_1})\, (a_{i_2 j_2 k_2})\, (a_{p' q r})\, (a_{p q' r'})\,.
\end{eqnarray}
Cancelling the common factor in the first and latter expression, which by hypothesis is different from zero, we obtain 
\be
(a_{p q r})\, (a_{p' q' r'})=(a_{p' q r})\, (a_{p q' r'})
\ee
as desired. Consequently, Eq.(\ref{stfour}) holds.

The extension of the previous result to any other set of indices is straightforward. E.g. if we enlarge the dimension of ${\cal H}_1$ up to 6, so that there two new possibilities: $i_5, i_6$, the whole argument holds for 
\be
(\mathds{1}-P_i)\vec{A}^{(i_1 i_2\ i'\ i'')\,(j_1 j_2 j_3 j_4)\,(k_1 k_2 k_3 k_4)}= 0\,,
\ee
with $(i'\ i')\in \{(i_3 i_5), (i_3 i_6), (i_4 i_5), (i_4 i_6), (i_5 i_6) \}$. This includes all the components of\\  $\vec{A}^{(i_1 i_2 i_3 i_4 i_5 i_6)\,(j_1 j_2 j_3 j_4)\,(k_1 k_2 k_3 k_4)}$, thus
\be
(\mathds{1}-P_i)\vec{A}^{(i_1 i_2 i_3 i_4 i_5 i_6)\,(j_1 j_2 j_3 j_4)\,(k_1 k_2 k_3 k_4)}= 0\,.
\label{stsix}
\ee
In a similar fashion the result is extended to arbitrary dimensions of the three subsystems.

\subsection{Extension to non-disjoint systems}

When considering non-disjoint systems we have to deal with collective indices, $I,J$ such that  $I\cap J\neq\emptyset$. Then, as was shown in section \ref{sec: triangular}, the triangular inequality for the square concurrences reads
\bea
C_{I\vartriangle J|\wideparen{I\vartriangle J}}^2 \leq 
C_{I|\wideparen{I}}^2+C_{J|\wideparen{J}}^2\,,
\label{general_triangular_square}
\eea
where $I\vartriangle J=I\cup J \setminus I\cap J$ is the set of indices in either $I$ or $J$, but not in both.

Following a procedure analogous to the one in previous subsections, it is straightforward to show 
that this relation becomes an equality {\em iff} either 
$C_{I|\wideparen{I}}$ or $C_{J|\wideparen{J}}$ are vanishing. More precisely, one starts by considering a tetrapartition $\{i,j,k,\wideparen{ijk}\}$, identifying $I=\{i,k\}$, $J=\{j,k\}$. Then Eq.(\ref{general_triangular_square}) becomes
\begin{equation}
C_{ij|\wideparen{ij}}^2\leq C_{ik|\wideparen{ik}}^2+C_{jk|\wideparen{jk}}^2\,.
    \label{AGen}
\end{equation}
From Eqs.(\ref{aGen}) the equality in this relation (\ref{AGen}) is equivalent to the condition
\be
-2\vec{A}^\dagger(\mathds{1}-P_i P_k)(\mathds{1}-P_j P_k)\vec{A}=0\iff(\mathds{1}-P_i P_k)(\mathds{1}-P_j P_k)\vec{A}=0\,.
\ee
So our goal is to prove the following implication
\begin{equation}
    (\mathds{1}-P_i P_k)(\mathds{1}-P_j P_k)\vec{A}=0\implies (\mathds{1}-P_iP_k)\vec{A}=0 \hbox{ or } (\mathds{1}-P_jP_k)\vec{A}=0\,.
\end{equation}
As for the disjoint case the strategy is to show ``by brute force" (using {\sf Singular}) that this is indeed the case for 4 qutrits, and then extend the result in a recursive way to higher dimensions. 
The procedure is completely analogous, but rather long and tedious, so we prefer to spare the reader.

\bibliographystyle{style2.bst}    
%
\bibliography{entanglement}	 

\end{document}